\documentstyle[preprint,revtex]{aps}                  
\def\ltap{\ \raisebox{-.5ex}{\rlap{$\sim$}} \raisebox{.4ex}{$<$}\ }
\def\gtap{\ \raisebox{-.5ex}{\rlap{$\sim$}} \raisebox{.4ex}{$>$}\ }

\newcommand{\be}{\begin{equation}}
\newcommand{\ee}{\end{equation}}
\newcommand{\beq}{\begin{eqnarray}}
\newcommand{\eeq}{\end{eqnarray}}

\begin{document}
\draft
\pagestyle{empty}                                      
\centerline{hep-ph/yymmnn}\vskip -.3cm
\centerline{\hfill  NTUTH-94-15}
\centerline{\hfill  NHCU-HEP-94-16}
\vskip 1cm
\begin{center}
\Large
Possibility of $m_t \ltap M_W$ for SUSY-type of Two Higgs Doublet Model
\end{center}
\vskip 1cm

\centerline{
Wei-Shu Hou and Ern-Bin Tsai}
\centerline{
Department of Physics, National Taiwan University,
Taipei, Taiwan 10764, R.O.C.} \vskip 1cm

\centerline{
Chao-Qiang Geng and Paul Turcotte} %
\centerline{
Department of Physics, National Tsing-Hua University,
Hsinchu, Taiwan 30043, R.O.C.}
\date{\today}

\begin{abstract}
Allowing for realistic uncertainties in input parameters,
we demonstrate that the present CLEO limit of $5.4\times 10^{-4}$
for inclusive $b\to s\gamma$ decay does not yet fully exclude
the $t\to bH^+$ decay possibility in supersymmetric type of
two Higgs doublet models.
Combined with direct search for $t\to bH^+$ via $H^+\to \tau^+\nu$
at the Tevatron, we conclude that $\tan\beta \sim 1$
is the ``allowed" window for $m_{H^+} < m_t < M_W$.
The possibility becomes excluded, however,
{\it if} the CLEO limit is pushed below $4\times 10^{-4}$.
\end{abstract}
{\vfill
}
%
\narrowtext
\pagestyle{plain}

\vskip 0.3cm
\noindent\underline{\it Introduction.} \hskip 0.5truecm
The CDF Collaboration at Fermilab has recently announced \cite{CDF}
the observation of dilepton and single lepton signals.
If confirmed, the conservative interpretation is the
production of $t\bar t$ pairs, with top quark
mass of order 174 GeV as suggested by event kinematics.
It has been pointed out \cite{Hou}, however, that these top-like signals
may in fact be due to {\it new} heavy quarks rather than the top quark.
The actual top quark,
defined to be the weak doublet partner of the bottom quark,
may have slipped through detailed scrutiny so far, and
could still be hiding below $M_W$ via scalar induced decays.
The reason is as follows.
If $m_t \ltap M_W$,
scalar induced {\it two body} decay modes may
dominate the top decay rate,
allowing the top quark to
evade earlier search efforts by suppressing
the semileptonic top decay branching ratio (BR)
into electrons or muons.
If such is the case, we have interesting physics
in store for us for next few years,
especially when LEP-II turns on \cite{Hou}:
a light top quark, plus scalars bosons that are even lighter,
{\it together with} new quarks that have mass of order weak scale.
The scenario is also appealing in the sense that it involves
heavy fermions (top) and scalar bosons in the picture at tree level.

Two possibilities are offered in ref. \cite{Hou}
for ``hidden truth" below $M_W$:
$t\to bH^+$ followed by $H^+\to c\bar s$ as dominant $H^+$ decay mode,
and flavor changing neutral coupling (FCNC) induced $t\to ch^0$ decay
($h^0$ is lightest neutral scalar) followed by $h^0 \to b\bar b$.
They are not disallowed by any known constraint.
The first possibility is realized \cite{ASW} in
a general class of two Higgs doublet model (2HDM),
usually called Model I, where
\begin{equation}
R_{\tau/c} \equiv {\Gamma(H\to c\bar s) \over \Gamma(H\to \tau\nu)}
\simeq {3\, m_c^2\over m_\tau^2},
\label{eq:RI}
\end{equation}
so BR$(H^+\to \tau^+\nu) \ltap 1/3$.
The second possibility is realized \cite{tch} in a generic 2HDM
that does not respect the Natural Flavor Conservation (NFC) condition
proposed by Glashow and Weinberg.
These two possibilities are rather interesting
in their own right, but appear to be somewhat exotic.
A third, more popular \cite{II} possibility, the so-called Model II
of 2HDM, and often referred to as the SUSY-type 2HDM,
seems to be ruled out by CDF via direct search \cite{tbH}
for $t\to b\tau^+\nu$,
and indirectly by the powerful CLEO bound \cite{CLEO}
on inclusive radiative decay,
\begin{equation}
{\rm BR}(b\to s\gamma) < 5.4 \times 10^{-4}.
\label{eq:bsA}
\end{equation}
It is often said \cite{Hewett} that the latter excludes
the possibility of $m_{H^+} < m_t$ altogether for the SUSY-type model.
In this Letter, however, we pay careful attention to details
and point out that $t\to bH^+$ and $m_t \ltap M_W$ are
in fact still allowed by {\it both} CDF's direct search
and CLEO's indirect constraint,
for a narrow  but interesting window in parameter space,
in the SUSY-type of two Higgs doublet model.
If the CLEO limit is improved to $4\times 10^{-4}$,
however, it would be sufficient to rule out the $t\to bH^+$
possibility beyond doubt for SUSY-type model.

\vskip 0.3cm
\noindent\underline{\it SUSY-type Model and CDF Direct Search.}
              \hskip 0.5 truecm
In the SUSY-type of two Higgs doublet model \cite{II},
which naturally occurs for minimal supersymmetric standard model (MSSM),
up-type quarks derive mass from
the vacuum expectation value (vev) of one scalar doublet,
while down-type quarks derive mass from the other.
The physical charged Higgs boson couples to fermions as
\begin{equation}
\frac{\sqrt{2}}{v}\, V_{ij}\,  \bar u_i \, \left( \cot\beta\,
    m_iL -  \tan\beta\, m_j R\right) d_j + h.c.,
\end{equation}
where $\tan\beta = v_2/v_1$ is the ratio of vev's
of the two Higgs doublet fields.
The usual prejudice is that $\tan\beta \sim m_t/m_b$,
and is therefore large. This leads to the expectation that
\begin{equation}
R_{\tau/c} \approx {m_\tau^2 \tan^2\beta \over  3\, (m_s^2 \tan^2\beta
                                                + m_c^2\cot^2\beta)},
\label{eq:RII}
\end{equation}
is dominated by $\tan^2\beta$ term and
therefore $H^+\to  \tau^+\nu$ is the dominant decay.
Note, however, that when $\tan\beta$ is small compared to $m_c/m_s$,
$R_{\tau/c}$ becomes very small and $H^+\to c\bar s$ is
the dominant mode.

The CDF collaboration has searched \cite{tbH} for $t\to bH^+$
in the domain
\begin{equation}
\left\{   \begin{array}{l}
              55\ {\rm GeV} < m_t < M_W + m_b, \\
              45\ {\rm GeV} < m_{H^+}, \\
              m_{H^+} + m_b < m_t,
             \end{array}   \right.
\label{eq:domain}
\end{equation}
via the $\tau\nu$ signature. From Fig.  3 of ref. \cite{tbH},
we see that for $BR(H \to \tau\nu) = 1$
they rule out almost the entire region.
The excluded region quickly diminishes, however,
as $B(H \to \tau\nu)$ drops.
For $BR(H \to \tau\nu) = 0.5$, the limit has all but disappeared.
Translating into $\tan\beta$ for the SUSY-type model,
one sees (Fig. 4 of ref. \cite{tbH}) that,
for
\begin{equation} \begin{array}{lccr}
\tan\beta \ltap 1,\ &\ \    &\ \
                     &\mbox{\rm (direct $t\to b\tau\nu$ search)},\cr
                 \end{array}
\label{eq:ub}
\end{equation}
the entire region of eq. (\ref{eq:domain}) is allowed,
but large $\tan\beta$ is ruled out.
Thus, the CDF result still allows for the
unconventional possibility that  $\tan\beta \ll m_t/m_b$.

\vskip 0.3cm
\noindent\underline{Implications of $b\to s\gamma$ Constraint.}
                  \hskip 0.5 truecm
Much interest \cite{Hewett2} has centered around
the 1993 CLEO limit \cite{CLEO} on inclusive $b\to s\gamma$ decay,
eq. (\ref{eq:bsA}).
It has been claimed \cite{Hewett} that this limit
rules out $m_{H^+} < m_t$ option altogether
in the SUSY-type 2HDM, so $t\to bH^+$ would be forbidden
by this constraint alone.
Subsequent work have emphasized that,
although the constraint is rather strong,
theoretical uncertainties \cite{Buras}
and effects of SUSY partners \cite{Borz}
may loosen the constraint such that
$t\to bH^+$ is not completely disallowed.
We will not discuss the effects of SUSY particles,
assuming that they are small
(or, we deal with a MSSM inspired scalar sector, without necessarily
invoking SUSY per se).
None of the above authors, however, have explored the
$m_t < M_W$ region, in large part because of blind faith in the old CDF
limit of $m_t > 91$ GeV. But we have argued in the introduction that
one has to be careful in this domain because of $t\to bH^+$ dominance over
$t\to bW^*$.
We now turn to making an analysis of $b\to s\gamma$
for light top possibility,
using state of the art QCD correction factors, but allowing for
theoretical and experimental errors,
in the spirit of Buras {\it et al.} \cite{Buras}.
Since details can be found more than abundantly in the literature,
we shall be concise in our treatment.

The standard technique in reducing uncertainties is to pin
BR$(b\to s\gamma)$ to inclusive semileptonic decay.
One has the formula,
\begin{eqnarray}
{\rm BR}(b\to s \gamma)& =& {\Gamma(b\to s\gamma)\over
\Gamma(b\to c e\bar{\nu})}\;{\rm BR}(b\to c e\bar{\nu})
\nonumber \\
&=& {\vert V_{ts}^*V_{tb}\vert^2 \over \vert V_{cb}\vert^2}
{6\alpha\over \pi f(m_c/m_b)}\;{
1\over \Omega (m_t/M_W,\mu)}
\;\vert C_7^{eff}(\mu)\vert^2
\;{\rm BR}(b\to c e\bar{\nu}),
\end{eqnarray}
where
\begin{eqnarray}
C^{eff}_7(\mu) &=& \eta^{{16\over 23}} C_7(M_W) +
{8\over 3} \left( \eta^{{14\over 23}} - \eta^{{16\over 23}}
\right) C_8(M_W)  + C_2(M_W) \sum_{i=1}^8 a_i \eta^{b_i},
\end{eqnarray}
summarizes short distance loop effects
induced by $W$ and $H^+$ at  scale $M_W$,
but brought down to the physical scale $\mu$
for the relevant $b\to s\gamma$ process,
by running the QCD coupling constant via
$\eta = \alpha_S(M_W)/\alpha_S(\mu)$.
The so-called ``large" QCD effect is exemplified by
the presence of  $C_2(M_W) = 1$
(for Standard Model and for 2HDM) term induced by operator mixing.
The phase space factor $f(z)$ is given by
$
f(z) = 1 - 8z^2 + 8z^6 - z^8 - 24z^4 \ln z.
$
The quantity $\Omega (x)$ contains the $O(\alpha_S)$ QCD corrections to the
semileptonic decay rate \cite{c24} and is given by 
\begin{equation}
\Omega (x, \mu)  \simeq  1-{2\alpha_S(\mu)\over 3\pi}
\left[\left( \pi^2 -{31\over 4} \right) (1-x)^2 + {3\over 2} \right].
\end{equation}
The scheme-independent numbers $a_i$ and $b_i$ are
given in ref. \cite{Ciu}.

In Model II (SUSY-type 2HDM), one finds \cite{HW,GW}
\begin{equation}
C_7(M_W) = -{1\over 2} A(x)
-B(h)-{1\over 6}\cot^2\beta\, A(h),
\label{eq:c7}
\end{equation}
where $x = m_t^2/M_W^2$, $h = m_t^2/m_{H^+}^2$, and
\begin{eqnarray}
A(x) &=& {-x\over 12(1-x)^4}
\left[6x(3x-2)\ln x+(1-x)(8x^2+5x-7)\right],
\nonumber\\
B(h) &=& {h\over 12(1-h)^3}
\left[(6h-4)\ln h+(1-h)(5h-3)\right].
\end{eqnarray}
Note that both the second and third terms of eq. (\ref{eq:c7})
are due to $H^+$ induced loop diagrams ($h$-dependent).
For third term, the $H^+$ coupling proportional to $m_t$ enters twice,
hence it has two powers of $\cot\beta$.
For second term,
the $H^+$ coupling proportional to $m_t$ and $m_b$ mutually balance each
other, such that no $\tan\beta$ dependence is left.
The reason that $b\to s\gamma$ is so powerful in constraining
SUSY-type Higgs sector is first due to the relative sign between
$W$ and $H^+$ effects, being always constructive,
and second because of this $\tan\beta$-{\it independent} term,
such that $H^+$ effect is always present even for small $\cot\beta$,
and is small only when $m_{H^+}$ is large compared to $m_t$.
This effect was originally pointed out by Hou and Willey \cite{HW},
and independently by Grinstein and Wise \cite{GW},
and has been utilized in the work of Hewett \cite{Hewett}
as the limits by CLEO sharpened in the past few years.

We wish to explore the allowed region of eq. (\ref{eq:domain})
as constrained by eq. (\ref{eq:bsA}),
but allowing for theoretical uncertainties
as advocated by Buras {\it et al.} \cite{Buras}.
Note that the region $m_t \ltap M_W$ is of interest here
for another reason:
When top is light, the SM expectation is lower than the present
CLEO inclusive bound of eq. (\ref{eq:bsA}),
hence enhancement due to $H^+$ effect may in fact be called for.
One could already see this if one extends the analysis of
Hewett, or that of Buras {\it et al.}, to below $m_t  < 91$ GeV.
The $m_t = m_{H^+} + m_b$ curve and
the curve depicting BR$(b\to s\gamma) < 5.4 \times 10^{-4}$ bound
tend to cross each other.
We illustrate this by plotting, in Fig. 1(a),
the effect of $b\to s\gamma$ constraint for fixed $\tan\beta$
in $m_{H^+}$--$m_t$ plane, by taking the central values of
the parameters,
\begin{equation}
\left\{ \begin{array}{l}
            m_c/m_b = 0.316 \pm 0.013, \\
            {\rm BR}(b\to ce\bar\nu) = 10.7 \pm 0.5 \% \\
            \alpha_S(M_Z) = 0.12 \pm 0.1 \\
            \mu = 2.5 - 10\ {\rm GeV},
           \end{array} \right.
\label{eq:param}
\end{equation}
and assume three generation KM unitarity.
For $\alpha_S(\mu)$, we run down from $\alpha_S(M_Z)$ using
the leading logarithmic approximation where
\begin{equation}
{\alpha_S(M_Z)\over \alpha_S(\mu)} = 1
          - \beta_0 {\alpha_S(M_Z)\over 2 \pi} \ln \left( {M_Z\over \mu}
\right) \label{v}
\end{equation}
with $\beta_0 = 11 - {2\over 3}N_f=23/3\ (N_f=5)$.
In Fig. 1, the solid curves from left to right are for
$\tan\beta = 0.6,\ 0.8,\ 1.0,\ 1.5,\ 3.0$ and $\infty$.
Note the convergence of curves for large $\tan\beta$.
The region of interest, $m_t > m_H + m_b$, is unshaded.
We then allow $2\sigma$ variation of parameters in eq. (\ref{eq:param}),
in particular, allowing $\mu$ to vary between $m_b/2$ and $2m_b$,
as well as an error range of $5\%$ for quark mixing angles.
For sake of simplicity, we do not consider $b\to s\gamma g$
(initial and final state gluon bremsstrahlung) as a
further correction \cite{AG} to be consistent in QCD.
The allowed region is now considerably expanded,
as shown in Fig. 1(b).
We show in Fig. 2
the allowed domain in $m_H{^+}/m_t - \tan\beta$ plane
for $m_t = $ 60, 70, 80 and 90 GeV.
In this parameter space, the distinction between
central value and $2\sigma$ error is not very significant,
and what is shown is for the latter.
For central value, the curves shift slightly to the right,
demanding a larger $\tan\beta$.
We conclude that $m_{H^+} \ltap m_t \ltap M_W$ is not
totally excluded by CLEO bound of eq. (\ref{eq:bsA}), if
\begin{equation}  \begin{array}{lccr}
\tan\beta \gtap 1,\ & \ \  & \ \
                      & \mbox{\rm (indirect $b\to s\gamma$ bound)}.\cr
                  \end{array}
\label{eq:lb}
\end{equation}
Smaller $\tan\beta$ leads to larger enhancements of $b\to s\gamma$ rate,
as is seen from eq. (\ref{eq:c7}), and is thereby excluded.

However, due to the convergence of curves for $\tan\beta > 1$,
and because of the proximity of the curves to the $m_t = m_H + m_b$ line,
the result of eq. (\ref{eq:lb}) is quite sensitive to
the limit on $B(b\to s\gamma)$.
We illustrate in Fig. 3 how the constraint changes
if the CLEO limit is pushed down to $4\times 10^{-4}$.
It is clear from Fig. 3(a) that, for central values,
$m_t > m_H + m_b$ is no longer possible.
With $2\sigma$ error tolerance,
we see that $\tan\beta > 1$ would be allowed by $b\to s\gamma$,
but it lies in the region ruled out by CDF, eq. (\ref{eq:ub}).
Hence, if BR$(b\to s\gamma) < 4\times 10^{-4}$,
combining with CDF bound, $t\to bH^+$ is no longer allowed in
SUSY type of two Higgs model.

\vskip 0.3cm
\noindent\underline{Discussion.} \hskip 0.5truecm
Combining eqs. (\ref{eq:ub}) and (\ref{eq:lb}),
we see that if $\tan\beta \sim 1$,
neither the direct search limit of CDF,
nor the present CLEO bound on inclusive $b\to s\gamma$,
strictly rules out the $m_{H^+} \ltap m_t \ltap M_W$ possibility.
The $b\to s\gamma$ decay can be slightly enhanced above SM expectation of
order 1-2$\times 10^{-4}$ for 55 GeV$\ltap m_t \ltap M_W + m_b$,
while $t\to bH^+$ could have allowed the top to evade CDF.
Note that $\tan\beta \sim 1$ is precisely the region of
parameter space where Model I and Model II
cannot be easily distinguished from studying top decay:
$H^+$ decay is dominated by $H^+\to c\bar s$,
and eq. (\ref{eq:RII}) merges with eq. (\ref{eq:RI}).
Fortunately, $b\to s\gamma$ sensitively
distinguishes between Model I and Model II,
another point that was originally stressed by Hou and Willey \cite{HW}.
If CLEO could improve their limit on $B(b\to s\gamma)$
down to $4\times 10^{-4}$
or below, the $t\to bH^+$ option would be ruled out for all $m_t$.
It is likely, of course, that CLEO would eventually yield a
number consistent with SM expectation of a few $\times 10^{-4}$,
and within error bars, our assertion would hold.

\acknowledgments

This work is supported in part
by the National Science Council of
Republic of China
under grants
NSC-83-0208-M-002-023 (W.S.H and E.B.T),
NSC-83-0208-M-007-118 (C.Q.G) and
NSC-83-0208-M-007-117 (P.T).

\newpage
\def\pl#1#2#3{
     {\it Phys.~Lett.~}{\bf #1B} (19#2) #3}
\def\zp#1#2#3{
     {\it Zeit.~Phys.~}{\bf #1} (19#2) #3}
\def\prl#1#2#3{
     {\it Phys.~Rev.~Lett.~}{\bf #1} (19#2) #3}
\def\pr#1#2#3{
     {\it Phys.~Rev.~ }{\bf D#1} (19#2) #3}
\def\np#1#2#3{
     {\it Nucl.~Phys.~}{\bf B#1} (19#2) #3}
\def\ib#1#2#3{
     {\it ibid.~}{\bf #1} (19#2) #3}

%


\vskip -1cm
\figure{Contours for $BR(b\to s\gamma) < 5.4 \times 10^{-4}$
        in $m_{H^+} - m_t$ plane for (left to right)
         $\tan\beta = 0.6,\ 0.8,\ 1.0,\ 1.5,\ 3.0$
         and $\infty$.
        Shaded region is for $m_t > m_{H^+} + m_b$.
        (a) is for central values of eq. (\ref{eq:param}),
        while (b) corresponds to $2\sigma$ error tolerance.}

\figure{Allowed region from Fig. 1(b) for
        $m_t = $60 (solid), 70 (dash), 80 (heavy dash) and 90 (dots) GeV.}

\figure{Same as Fig. 1 except for $BR(b\to s\gamma) < 4 \times 10^{-4}$.}


\begin{references}
%
\vskip-1cm
%
\bibitem{CDF} CDF Collaboration, F. Abe et al.,
{\it FERMILAB-PUB-94/097-E}.

%
\bibitem{Hou} W. S. Hou,
      {\it Phys. Rev. Lett.} {\bf 72}, 3945 (1994).
%
\bibitem{ASW}
L.~F.~Abott, P.~Sikivie and M.~B.~Wise, \pr{21}{80}{1393}.

%
\bibitem{tch} W. S. Hou, {\it Phys. Lett.} {\bf B296} (1992) 179;
              L. J. Hall and S. Weinberg, \pr{48}{93}{R979}.
%
\bibitem{II}
Cf. J. F. Gunion $et\ al.,$ {\sl The Higgs Hunter's Guide},
Addition-Wesley Pub., 1990.
%
\bibitem{tbH}
CDF Collaboration, F. Abe {\it et al.},
{\it Phys. Rev. Lett.} {\bf 72} (1994) 1977.
%
\bibitem{CLEO}
CLEO Collaboration, R.~Ammar et. al.,
{\em Phys. Rev. Lett.} {\bf 71} (1993) 674.
%
\bibitem{Hewett}
J.~L.~Hewett, \prl{70}{93}{1045}.
%
\bibitem{Hewett2}
For a recent review, see
J. L. Hewett, {\it SLAC-PUB-6521}, and references therein.
%
\bibitem{Buras}
A. J. Buras, $et\ al.,$ {\it MPI-Ph/93-77}, 
and references therein.
%
\bibitem{Borz} See, {\it e.g.} F. M. Borzumati,
{\it Print-93-0025},
and references therein.
%
\bibitem{c24} N. Cabibbo, G. Corb\`o and L. Maiani, {\it Nucl. Phys.}
{\bf B155} (1979) 93;
%
M. Jezabek and J.M. K\"uhn, {\it Nucl. Phys.}  {\bf B320} (1989)
20 and   {\bf B314} (1989) 1.
%
%
\bibitem{Ciu}
M. Ciuchini. $et\ al.,$ {\it Phys. Lett.} {\bf B316} (1993) 127.
%
\bibitem{HW}
W. S. Hou and R. S. Willey, {\it Phys. Lett.} {\bf B20
2} (1988) 591; {\it Nucl. Phys.} {\bf B326} (1989) 54.
%
\bibitem{GW}
B. Grinstein and M. B. Wise, {\it Phys. Lett.} {\bf B201} (1988) 274.
%
\bibitem{AG}
A. Ali and Ch. Greub, {\it Z. Phys.} {\bf C60} (1993) 433.
\end{references}
\end{document}